\documentclass[]{aa}
\usepackage{graphicx}
\input{psfig}
\usepackage{txfonts}
\begin{document}

\title{Sensitivity of low degree oscillations to the change in solar abundances}
\author{
 A.~Zaatri \inst{1,}\inst{2}
\and  J.~Provost \inst{1}
\and G.~Berthomieu\inst{1}
\and P.~Morel\inst{1} 
\and T. Corbard\inst{1} }
\institute{D\'epartement Cassiop\'ee, UMR CNRS 6202, Observatoire de
la C\^ote d'Azur, BP 4229, 06304 Nice CEDEX 4, France 
\and Centre de Recherche en Astronomie, Astrophysique et G\'eophysique, BP 63,
Route de l'Observatoire, Bouzar\'eah, 16340, Alger, Alg\'erie}
\titlerunning{Sensitivity of low degree oscillations}
\authorrunning{Zaatri et al.}
\date{Received date / Accepted date}

\abstract
{The most recent determination of the solar chemical composition,
  using a time-dependent, 3D
hydrodynamical model of the solar atmosphere, exhibits a
significant decrease of C, N, O abundances compared to their
previous values. 
Solar models that use these new abundances are not
consistent with helioseismological determinations of the sound speed
profile, the surface helium abundance and the convection zone depth.} 
{We investigate the effect of changes of
solar abundances on low degree p-mode and g-mode characteristics which are strong constraints of the solar core. We consider  
particularly the increase of neon abundance in the new solar mixture in order to reduce the discrepancy between models using new abundances and helioseismology.}
{The observational determinations of solar frequencies from the GOLF instrument are used to test solar models computed with different chemical compositions.
We consider in particular the normalized small frequency spacings in
the low degree p-mode frequency range.}
{Low-degree small frequency spacings are very sensitive to changes in the heavy-element abundances, notably neon. 
We show that by considering all the seismic constraints, including the
small frequency spacings, a rather large increase of neon abundance by
about ($0.5\pm0.05$)dex can  be a good solution to the discrepancy between solar models that use new
abundances and low degree
helioseismology, subject to adjusting slightly the solar age and the
highest abundances. We also show that the change in solar abundances, notably neon, considerably affects g-mode frequencies, with   
relative frequency differences between the old and the new models
higher than 1.5\%.} 
{}
\maketitle

\section{Introduction}
The precise measure of characteristics of the observed p-modes has been used to probe most 
of the layers inside the sun. For example, seismic sound speed and density determinations can be used to constrain the interior of the sun
anywhere except at the surface and in the core. Helioseismology also constrains to the surface helium abundance and the depth of the
convection zone. 
 However, the  small number of p-modes (only low
degree p-modes) able to reach the solar core is not sufficient
to probe this region using inversion techniques. The solar core is crossed by
thousands of g-modes, able to bring much information
from this region. The g-modes have not yet been unambiguously identified because of their evanescent nature through the convection zone and low amplitude at the surface 
but ongoing work is devoted to try to extract them for the existing
long time series of SOHO data and to propose new observational and
strategies to detect them. 

 New determinations of solar heavy element abundances using a 3D, NLTE
analysis of the solar spectrum has been provided by Asplund et
al. (2005; AGS). Previous 1D, LTE determinations
are available (Grevesse and Noels~1993- GN; Grevesse and  Sauval~1998 - GS).
Relative to GN abundances, the new AGS abundances are, among others, lower
in $C$, $N$, $O$, $Ne$ elements by respectively  $0.16$~dex,
$0.19$~dex,  $0.21$~dex and $0.24$~dex.
Consequently,
the new chemical determination gives a smaller metallicity $Z X$
compared to the older ones.
The new determination of solar elements is more accurate than the
older one (Grevesse et al. 2005), but it has been shown that it leads to 
solar models that disagree with the helioseismological determinations of solar internal structure parameters (e.g. Turck-Chi\`eze et 
al. 2004; Bahcall et al. 2005; Guzik et al. 2005).

In this paper we study the sensitivity of the solar core properties, through low degree p-modes
and g-modes, to the change of solar mixture. 
The chemical solar abundances that we used are those of
Grevesse \& Noels~(1993),  Grevesse \& Sauval (1998) and Asplund et al. (2005). The other mixtures that we chose
include the solar abundances of Asplund et al (2005) changing mainly the neon
abundance. This set of solar mixtures allows us to study the influence of the neon abundance on small frequency separations and to test the
possibility of improving the accordance between models that use new abundances and helioseismic observations (Antia et Basu 2005; Bahcall et al. 2005).
This is indeed possible because neon photospheric abundance cannot be determined directly 
due to the lack of suitable absorption lines in the solar spectrum 
induced by the noble gas nature of neon. The estimations  
of the solar Ne abundance are still controversial 
(Drake \& Tesla 2005, Young 2005, Schmelz 2005).
 Here we analyze how the models fit both the global constraints 
(seismic sound speed, surface helium abundance and convection zone depth) and the small 
frequency separations in the low degree p-mode frequency range. 
We use the  determination of these  mode frequencies obtained from 
the GOLF experiment by Gelly et al. (2002) and more recently  by Lazrek 
et al. (2007) who have corrected these frequencies for the solar cycle effect.
 The sensitivity of gravity modes to the new abundances have also  been 
estimated. Preliminary results of this work have been presented by Zaatri et al. (2006).

\begin{centering}
\begin{table}
\begin{tabular} [0.5pts]{p{0.9cm} p{0.75cm} p{0.75cm} p{0.75cm} p{0.75cm} p{0.75cm}
p{0.75cm}}
\hline
  ~~\\
  ~&\tiny{$A(Ne)$}&\tiny{$(Z/X)_S$}&\tiny{$Y_S$}&\tiny{$r_{ZC}$}&\tiny{$T_c^{7}$}&\tiny{$P_0$}\\
   ~&~&~&~&~&~&\\
  \tiny{M-GN}&8.08&0.0245&0.2437&0.7133&1.574&35.08\\
  \tiny{M-GS}&8.08&0.0232&0.2462&0.7165&1.574& 35.03\\
  \tiny{M-GS$^*$}&8.08&0.0231&0.2429&0.7153&1.571&35.15\\
  \tiny{M-AGS}&  7.84& 0.0166&0.2279&0.7292&1.549&35.68 \\
  \tiny{M3}& 8.10& 0.0179& 0.2329& 0.7237& 1.555& 35.49 \\
  \tiny{M4}&8.29&0.0192& 0.2381& 0.7181&  1.559& 35.29 \\
  \tiny{M5}&8.35& 0.0198& 0.2402&  0.7162&  1.561&35.21  \\
  \tiny{M6}&8.40& 0.0203 &0.2417 & 0.7144  & 1.563 &35.15 \\                   
  \tiny{M7}&8.47&0.0212&0.2445&0.7119&1.565&35.04\\
  \tiny{M8}&8.35& 0.0213&0.2439&0.7142&1.566&35.05\\
  
  \hline
 \end{tabular}
\caption{Global characteristics of the computed solar models.
$A(Ne)$ is the neon abundance in dex, $(Z/X)_S$ is the surface
 metallicity, $T_c^{7}$= $T_c*10^{-7}$, $T_c$ the central temperature
 in Kelvin. $P_0$ is the characteristic period (in minutes)
of low degree gravity modes. The different models are computed with the following solar
 abundances: M-GN: GN; M-GS: GS;
M-GS$^*$: GS with sulfur abundance of GN;
M-AGS: AGS; M3, M4, M5, M6, M7: AGS with the indicated change of
 the neon abundance; M8: in addition to the change of neon in AGS, 
C, N, O, Mg and Si have been increased until the maximum of their error bars (see Asplund et
al. 2005) and Ar by 0.40 dex.}\label{}
\end{table}
\end{centering}

\section{Solar models with new abundances}
We have computed solar models with different sets of
heavy element abundances by using the stellar evolution code CESAM (Morel 1997).
 We use OPAL opacity tables\footnote{http://www-pat.llnl.gov/Research/OPAL/},
calculated for each mixture, and Alexander and Ferguson opacity tables at low temperatures ($T<6000K$). Nuclear
reaction rates are from NACRE compilation ( Angulo et al.  1999) and equation of state tables are those of OPAL  (Iglesias \& Rogers 1991). We assume the convection treatment given
by Canuto and Mazitelli (1991). All the models include the microscopic 
diffusion of the chemical elements according to the Michaud \& Proffitt 
(1991) description. Models are calibrated for a solar age $t$=4.6~Gyr at the solar
radius, the solar luminosity ($R_{\odot}= 6.9599\times10^{10}$~cm,
$L_{\odot}= 3.846\times10^{33}$~erg/s, Christensen-Dalsgaard et al. (1996)
) and the solar surface metallicity $Z/X$ of the various mixtures.

\begin{figure}[htbp]
\centering
\small
\includegraphics[width=8.5cm]{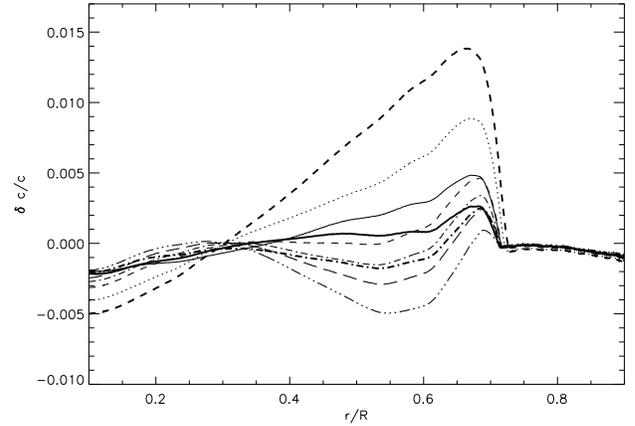}\hfill
  \caption{Relative sound speed differences between the Sun and the models: 
M-GN (heavy full), M-GS (full), M-AGS (heavy dashed), M3 (dotted), M4 (short-dashed), 
M5 (dashed-dotted),M6 (long-dashed),M7 (dashed-3*dotted), M8 (heavy dashed-dotted).
}\label{dcsc}
\end{figure}

\begin{figure}[htbp]
\centering
\small
\includegraphics[width=8.5cm]{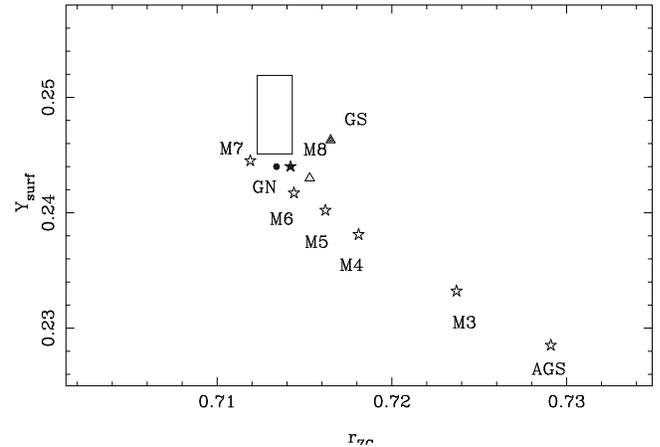}
\caption{Characteristics of the solar envelope,  surface helium abundance ($Y_S$) and the location of the base of the convection zone ($r_{ZC}$),
for the models: M-GN (filled circle), M-GS (filled triangle),
M-GS$^*$ (open triangle), M-AGS (open star), M3 to M7 which is a set of 
models that use AGS mixture by varying its neon abundance (open star) and M8 (filled star). 
The box represents the seismic values with their errors
(Basu and Antia 2004).}\label{ys-zc}
\end{figure}

\begin{figure*}
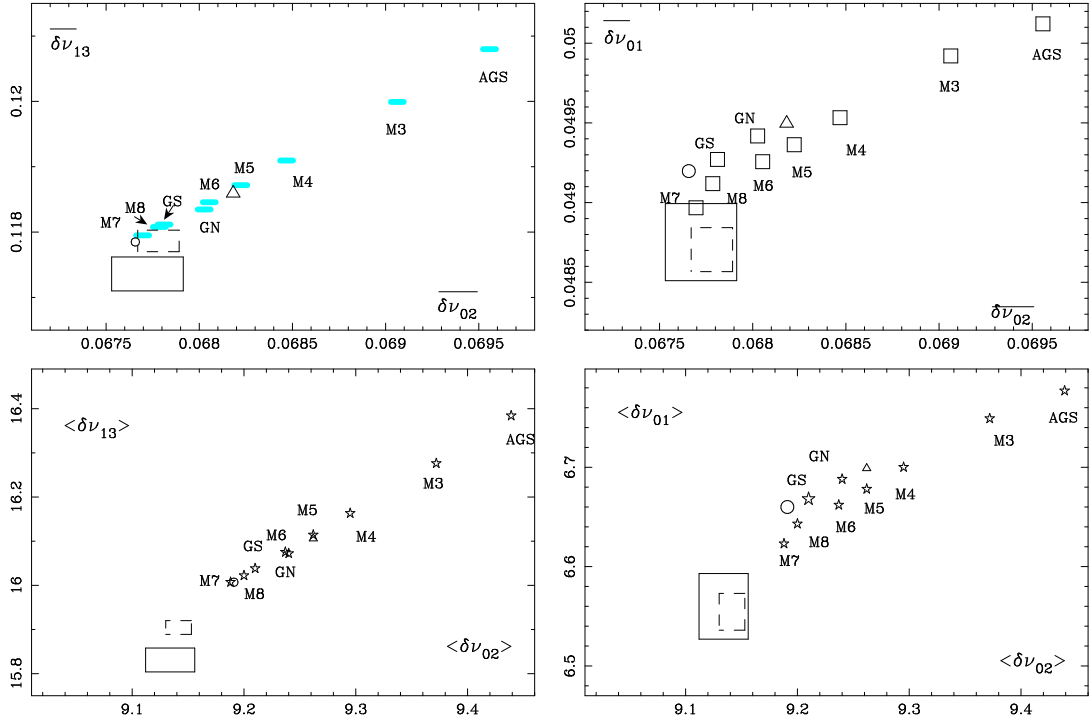

\centerline{
\small
\vbox{
{\hbox{
\includegraphics[width=7cm]{fig_dnu02-13.ps.aa.norm+}
\hspace{0.2cm}
\includegraphics[width=7cm]{fig_dnu02-01.ps.aa.norm+}
}}
\vspace{0.2cm}
{\hbox{
\includegraphics[width=7cm]{fig_dnu02-13.ps.aa+}
\hspace{0.2cm}
\includegraphics[width=7cm]{fig_dnu02-01.ps.aa+}
}}
}
}
 \caption{Upper panel left: Renormalized mean frequency small spacing
 $\overline{\delta\nu}_{13}$
as a function of the renormalized mean frequency small spacing $\overline{\delta\nu}_{02}$ for the
different models compared to GOLF observations (full box Gelly et
al. 2002, dashed  box Lazrek et al. 2007). Upper panel right : same for renormalized mean frequency small spacing  $\overline{\delta\nu}_{01} $ as a
function of the  renormalized mean frequency small spacing
$\overline{\delta\nu}_{02}$. Lower panel: for comparison  the same quantities are given before renormalization (in $\mu$Hz), namely $<{\delta\nu}_{13}>$ and  
$<{\delta\nu}_{01}>$ as a function of $<{\delta\nu}_{02}>$. The small circle corresponds to a M-GN model calibrated at 4.65Gyr,
the open triangles correspond to M-GS$^*$
model. }\label{spacings}
\end{figure*}

Table 1 summarizes the characteristics of the
solar models at both the surface and the core and their chemical composition is given. The surface helium abundance and the location of the 
base of the convection zone $Y_s$ and $r_{ZC}$ are to be 
compared to their seismic determinations $Ys=0.2485\pm0.0034$, $r_{ZC}=(0.7133\pm0.001)R_{\odot}$ by Basu \& Antia (2004). These authors have demonstrated 
that these seismic determinations are not sensitive to the change in solar abundances.\\

Figure \ref{dcsc} shows relative differences between seismic sound speed and the one
determined from our different solar models. Figure \ref{ys-zc} shows a
comparison between $Y_s$ and $r_{ZC}$ values of the computed models and their seismic
determinations. The worse concordance between the model using Asplund et al. abundances and the
seismic model is shown by a relative difference in sound speed that peaks at
1.5\% just below the convection zone. The surface helium abundance and the location of the base of the convection zone are also very far from their seismic values. 

Models M3, M4, M5, M6 and M7 give an idea of how large the neon abundance increase has to be in order to reduce this discrepancy. In all these models the neon has 
been pushed out of its error bar ($\pm0.06$dex). Before looking for an optimal value of neon, we notice that the larger the neon abundance, the larger the surface
helium abundance, the larger the convection zone depth and the higher the sound speed.      
The augmentation of neon in M4 makes the model's sound speed close to the seismic profile but keeps $Y_s$ and $r_{ZC}$ far from their seismic box, 
even if they become better than the M-AGS ones.
A slightly higher increase of neon (e.g. M7) improves the surface envelope characteristics ($Y_s$, $r_{ZC}$) but makes the 
sound speed much higher than the seismic one. Therefore, a compromise has to be found for the neon abundance value to satisfy $Y_s$, $r_{ZC}$ and sound speed constraints. We estimate the neon abundance to be $8.39\pm0.05$dex, subject to some remaining differences between the models and the observation. 
First, the lowest values of this range lead to models that have $Y_S$ 
and $r_{ZC}$ values about $2\sigma$ far from their seismic values. Second, its highest values infer models with relative differences between their sound speed profile and 
the seismic one that are just three times lower than the relative difference between M-AGS and seismic sound speed profiles at their
peaks.

In order to see the influence of the other heavy elements abundances on the change of the considered model characteristics ($Y_S$, $r_{ZC}$ and sound speed) we constructed the M8 model.
This model has a neon abundance that is situated at the bottom of our given neon increase interval (8.35dex) and has C, N, O, Si and Mg abundances that are increased until the maximum
of their error bars (see Asplund et al. 2005). The abundance of argon has also been increased by 0.4dex, as this element is another noble gas of the solar mixture. We deduce that the sound speed of the M8
model does not change much compared to the model that has the same increase of neon (M5), except at the deeper layers of the sun. However, the values of $Y_S$ and $r_{ZC}$ become closer
to their seismic values compared to those of the M5 model. This means
that the increase of other heavy element abundances inside their error
bars  simultaneously improves the agreement between the three
considered seismic determinations and those of the models. This can
lead to a reduction of the supposed values of neon as they give good agreement in sound
speeds to ($8.34\pm0.05$)dex. For this range, relative differences between seismic and theoretical $Y_S$ and $r_{ZC}$ still have a maximum of $2\sigma$.

For comparison we have considered the models M-GN and M-GS  with the 
previous mixtures GN  and GS. The model  M-GN has a convective zone depth
 close to the seismic one but a too small surface helium abundance.
On the contrary, model  M-GS has a good surface helium content but a thinner 
convective zone. Since in Figure 2, the position of the M-GS model does not follow
the general trend of the other models, we have looked  in more detail at the 
differences between the two mixtures. We noted that the sulfur abundance of
GS mixture (7.33  dex) is larger than  the GN value (7.21  dex),
due to improved oscillator strengths (Biemont et al. 1993), and than the AGS value (7.14 dex).
 We computed a model 
M-GS$^*$   with the GS mixture but with GN sulfur abundance.
It appears that such a variation of sulfur  notably lowers the surface 
helium abundance of the model.
         
\begin{figure*}
\centering
\small
 \includegraphics[width=11cm]{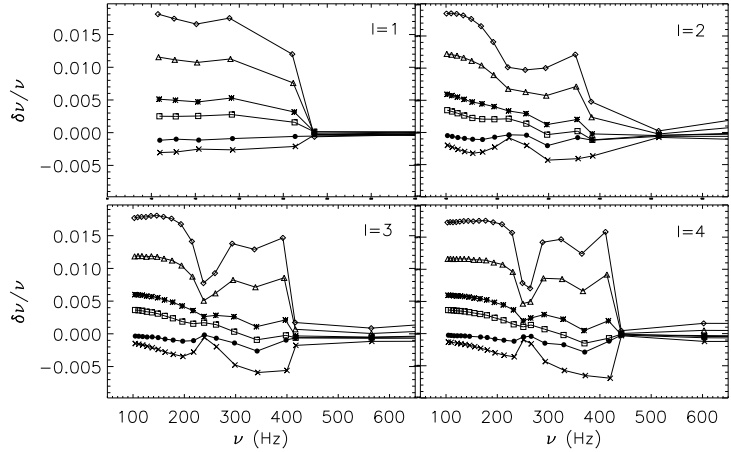}
  \caption{Relative frequency differences between the gravity modes and the first low frequency p-modes of 
M-GN and M-AGS ($\diamond$), M3 ($\triangle$),
M4($\ast$) , M5 ($\square$), M7($\times$), M8($\bullet$).
}\label{diffnu}
\end{figure*}

\section{ Fit to the low-degree small frequency spacing constraints}

Small low degree frequency spacings are a well known diagnostic of the solar core.
In order to compare our theoretical results to the observational ones, we use the latest results of Lazrek
et al. (2007) and those of Gelly et al. (2002) in the measurement
of low degree solar frequencies from the GOLF experiment. We examine the small frequency spacings $\delta\nu_{02}$,
$\delta\nu_{13}$ and $\delta\nu_{01}$ which are combinations of
acoustic modes penetrating differently towards the center and that are
thus very sensitive to this region.

$$\delta\nu_{02}(n)= \nu_{n,\ell=0} - \nu_{n-1,\ell=2}, $$
$$\delta\nu_{13}(n)= \nu_{n,\ell=1} - \nu_{n-1,\ell=3}, $$
$$\delta\nu_{01}(n)=2 \nu_{n,\ell=0} - (\nu_{n,\ell=1} + \nu_{n-1,\ell=1}).$$

However, the small spacings are slightly dependent on the solar atmosphere which is highly simplified in the solar models.
Roxburgh \& Vorontsov (2003) have demonstrated that the ratio of the small to large separations of acoustic oscillations is essentially independent of the structure of the outer layers. 
We thus renormalize the small spacings by considering
 the ratios $\delta\nu_{02}(n)/\Delta\nu (n,\ell=1)$,
 $\delta\nu_{13}(n)/\Delta\nu (n+1,\ell=0)$ and  $\delta\nu_{01}(n)/\Delta\nu (n,\ell=1)$ where the large separation is given by:

$$\Delta\nu(n,\ell)= \nu_{n,\ell} - \nu_{n-1,\ell} $$

We then compute both for our models and for the observations the
mean of these renormalized small frequency spacings
 $\overline{\delta\nu_{02}}$,
$\overline{\delta\nu_{13}}$ and $\overline{\delta\nu_{01}}$ for
radial orders from 16 to 24. This corresponds to a frequency
range about 2500 -- 3600~$\mu$Hz. The low limit of this range
ensures that  the behavior of the frequency is almost asymptotic and
the high limit corresponds to observed modes with very high
accuracy. For higher frequencies, the accuracy decreases rapidly. 
Figure \ref{spacings} gives both renormalized and non renormalized 
mean small spacings for the models of Table 1 and for the observations.
 The dimensions of the symbols in the  upper panel of Figure \ref{spacings} reflect the uncertainty on the plotted quantities corresponding to an uncertainty of $0.01\mu$Hz for the
 numerical frequencies.  
It shows that the renormalization gives results closer to the 
observations because it eliminates the differences 
between surface properties of the models and the sun. 
The remaining discrepancies are due to differences in the 
structure of the deep solar interior.\\ 
As expected, Figure \ref{spacings} shows that the small frequency spacings 
of the M-AGS model are far from the observations, contrary to those of the M-GN 
and M-GS models. 
We note that the M-GS model is closer to the observations than the M-GN model,
contrary to the M-GS$^*$ model. 
The small frequency spacings of the models that use an AGS mixture
with different values of the neon abundance decrease as the neon increases in almost a regular way. 
The model M7, which uses the highest value of neon abundance in our considered set of models, is shown to have the best agreement between its small spacings and 
 the observations. However, this model has a much higher sound speed than the seismic one (see Figure\ref{dcsc}), which makes it a less acceptable model. We
 also show that a slight increase of some other heavy elements has an effect on the change of small frequency spacings as well. We find that for the model M8 the three
 considered renormalized small spacings are closer to their observational determinations than those of the model M5.           
  
However the small spacings are also
sensitive to the solar age, due to the change in time of the mean 
molecular weight in the nuclear core.

For example, Morel et al. (1997) showed that an increase in age of 100 Myr gives a decrease of $0.1\mu$Hz
of both 
$\overline{\delta\nu_{02}}$ and $\overline{\delta\nu_{13}}$,
with a small relative increase of sound speed of around 10$^{-3}$
and a slight increase of the thickness of the convection zone
of 0.002$R\odot$ and no noticeable change of the surface helium abundance. 

We added, in Figure \ref{spacings} a GN model which is calibrated at 4.65 Gyr to see 
the influence of the solar age on small frequency spacings. \\
After showing the influence of solar abundances on low degree small frequency spacings, we still believe that in order to resolve the new abundances, 
a compromise between the neon abundance, the small frequency spacings
and the constraints of the preceding paragraph can be reached by
slightly adjusting some heavy elements inside their error bars and the age of the sun. Our
suggested value of neon ($8.34\pm0.05$) is confirmed  by considering the low degree small
frequency spacing constraint.        

\section{Gravity modes}
Adiabatic frequencies of the models have also been computed for 
the range from 100 to 800~$\mu$Hz and for low degrees ($0 <\ell < 6$) including both g-modes and mixed modes. 
The periods of low frequency gravity modes are
proportional to the characteristic period $P_0$ which is given in Table 1
($P_0={2\pi^2}/{\int_0^{r_{ZC}}{({N}/{r})dr}}$, where N is the 
Brunt-V\"aiss\"al\"a frequency). The lowest $P_0$ difference between M-GN and the
other models is obtained for M8, leading to the closest g-mode frequencies.
The frequency differences between the M-GN and the other models
are given in  Figure \ref{diffnu}. 
We note that the g-modes are more influenced by the change of abundances
than the low frequency p-modes.
As expected the  value of $\delta\nu/\nu$ at very low frequency
is close to its asymptotic value $\delta P_0/ P_0$. The biggest frequency difference  is obtained for M-AGS for which low g-mode frequencies are 1.5\% lower. This
difference 
has a minimum for all the models around $250\mu$Hz. It has been shown that the g-modes around this frequency have 
a mixed character (Provost et al. 2000) and  are sensitive to both the
sound speed and the Brunt-V\"aiss\"al\"a frequency variations.
Their frequencies do not depend much on any change in the models. 
The lowest difference in the frequencies compared
to M-GN is obtained for M8 which is expected since they have very similar structure parameters.\\

\section{Discussion and conclusion}

We study the sensitivity of low degree frequency spacings to the change on solar heavy element abundances.
We constructed several solar 
models with different solar mixtures. The spacings are considered as helioseismic constraints of the solar core as they are very sensitive to 
this deep region. Therefore, they are used to test the reliability of the 
solar models in addition to the envelope constraints (convection zone depth 
and helium surface abundance) and the sound speed profile.      
Surface effects have been removed from these spacings by using a 
renormalization prescribed by Roxburgh and Vorontsov (2003). 
Their observational values have been taken from the recent GOLF measurements 
(Lazrek et al. 2007; Gelly et al. 2002).\\
 We found that low degree small frequency spacings are very sensitive to the 
metallicity of the   models. The mean spacings of a model  using 
Asplund et al.(2005) abundances are much higher than the ones of 
a model using Grevesse and Noels (1993) values and are incompatible
with those determined from 
the GOLF observations. Similar results were found 
by Basu et al. (2007) by comparing  BISON observations of low degree 
solar oscillations with models 
using different abundances and numerical codes. They conclude that 
low surface metallicity models can be ruled out.
 These two studies strengthen 
the fact that new solar abundances lead to solar 
models which disagree with helioseismology measurement 
in the core as well as in the other regions of the 
solar interior.

We confirm, as was also mentioned by 
several authors, that the sound speed profile, convection zone depth and 
 helium surface abundance of a model using the revised abundances are 
also far from their helioseismic determinations, unlike the ones
 of a model using the old abundances.\\ In addition to these two main models we constructed five other models that use new solar
 abundances with a significant change of the neon abundance. This has been done following the Antia \& Basu (2005) suggestion to resolve the new 
 abundances. We found that the small spacings are very sensitive to the neon abundance value and decrease almost regularly when the neon increases. 
The discrepancy between models and observations is reduced simultaneously for the small frequency spacings and the 
 other constraints when the neon abundance is considerably increased. However, the neon abundance that gives the best agreement between the models and the helioseismic
 determinations is hard to reach as it is a compromise solution between all these quantities. For instance, a model using a neon value increased by 0.45dex (M4) makes its
  sound speed very close to the seismic sound speed but keeps its envelope characteristics (convection zone depth and helium surface abundance) far from their observational
  values. A model using a neon abundance increased by 0.63dex (M7) has
  the opposite effect.\\ 

As expected, the search for the neon abundance value that gives a  good agreement between models using new abundances and the seismic constraints 
including the small frequency spacings becomes easier if C, N, O, Si, Mg and Ar abundances are also slightly increased.
 Other elements may also play a significant role. We show for example
 that an increase of sulfur abundance, as is the case for the GS
 mixture, noticeably increases the surface helium abundance and lowers the small frequency spacings.Also, the solar age is a crucial feature in the determination of low-degree frequencies. 
Indeed, we tested a model using old abundances with an age 50 Myr older than the age we have used to compute all the models (4.6Gyr) and found that this
change brings the spacings closer to the observations. 

In conclusion, we show that,if the new solar
mixture of Asplund et al. (2005) is confirmed, an increase of the neon abundance by ($0.5\pm0.05$)dex can resolve the current disageement caused by this mixture, subject to adjusting slightly the highest
heavy-element abundances and the age of the sun in order to satisfy all the seismic constraints, notably the low-degree small frequency spacings. 
 Our estimated increase of neon is slightly 
higher than that of Bahcall et al.(2005), which is $0.45\pm0.05$. 
 However, in this work only low degree modes are
  used, while the higher degree modes also provide independent
  constraints. Also our mode,l which is the closest to 
the observations, has a surface metallicity content $Z$ smaller than  
the value determined by Antia and Basu (2006) from higher degree modes 
using the dimensionless sound speed derivative in the solar convection  zone.

Our last investigation in this work has been the calculation of g-mode
frequencies since the detection of g-modes is one of the current challenges of solar observers. 
As expected, the solar model using new abundances has the highest frequency differences to the model using old abundances, which go up to 4 $\mu$Hz. 
We show that modes with frequencies around $250\mu$Hz and degrees larger than 2 are less sensitive modes to the change in the abundances, 
with differences less than 2 $\mu$Hz.\\

{\bf Acknowledgments}:
We thank B. Pichon for his technical help, D.R. Alexander for giving us low temperature opacity tables for the revised mixture
and the OPAL group for their online opacity tables code. 
We are grateful to G. Grec and M. Lazrek for communicating their results in advance of publication and to H.M. Antia for his constructive remarks.
We also thank the ``Programme Pluri-Formations Ast\'erosismologie'' from OCA for the financial support.

\begin{small}

\end{small}
\end{document}